\documentclass[pdflatex,sn-mathphys-num]{sn-jnl}

\usepackage{epsfig}
\usepackage{comment}
\usepackage{subfigure}
\usepackage{caption}
\usepackage[section]{placeins}
\usepackage{graphicx}
\usepackage[notrig]{physics}
\usepackage{units}
\usepackage{esint}
\usepackage{natbib}
\usepackage{xcolor}

\usepackage{multirow}
\usepackage{amsmath,amssymb,amsfonts}
\usepackage{amsthm}

\usepackage[title]{appendix}
\usepackage{xcolor}
\usepackage{textcomp}
\usepackage{manyfoot}
\usepackage{booktabs}
\usepackage{algorithm}
\usepackage{algorithmicx}
\usepackage{algpseudocode}
\usepackage{listings}

\begin{document}

\title[Adhesive-joint lifetime prediction]{\bf Multiscale analysis and lifetime prediction of adhesive lap joints in contact with aggressive environments}

\author*[1]{\fnm{M.~P.} \sur{Ariza}}\email{mpariza@us.es}

\author[2,3]{\fnm{M.} \sur{Ortiz}}\email{ortiz@caltech.edu}

\affil*[1]{\orgdiv{Escuela Técnica Superior de Ingeniería}, \orgname{Universidad de Sevilla}, \orgaddress{\street{Camino de los descubrimientos, s.n.}, \city{Sevilla}, \postcode{41092}, \country{Spain}}}

\affil[2]{\orgdiv{Division of Engineering and Applied Science}, \orgname{California Institute of Technology}, \orgaddress{\street{1200 E.~California Blvd.}, \city{Pasadena}, \postcode{91125}, \state{CA}, \country{USA}}}

\affil[3]{\orgdiv{Centre Internacional de Métodes Numerics en Enginyeria (CIMNE)}, \orgname{Universitat~Politècnica~de~Catalunya}, \orgaddress{\street{Jordi~Girona~1}, \city{Barcelona}, \postcode{08034}, \country{Spain}}}

\abstract{We formulate a multiscale model of adhesive layers undergoing impurity-dependent cohesive fracture. The model contemplates three scales: i) at the atomic scale, fracture is controlled by interatomic separation and the thermodynamics of separation depends on temperature and impurity concentration; ii) the mesoscale is characterized by the collective response of a large number of interatomic planes across the adhesive layer, resulting in a thickness-dependence strength; in addition, impurities are uptaken from the environment and diffuse through the adhesive layer; and iii) at the macroscale, we focus on lap joints under the action of static loads and aggressive environments. Within this scenario, we obtain closed form analytical solutions for: the dependence of the adhesive layer strength on thickness; the length of the edge cracks, if any, as a function of time; the lifetime of the joint; and the dependence of the strength of the joint on time of preexposure to the environment. Overall, the theory is found to capture well the experimentally observed trends. Finally, we discuss how the model can be characterized on the basis of atomistic calculations, which opens the way for the systematic exploration of new material specifications.}


\maketitle

%\tableofcontents

\section{Introduction}\label{sec:1}

Hybrid bonded structures, where steel components are adhesively bonded to polymeric or composite components, offer significant advantages in ship-building design, including attractive strength-to-weight ratios, and can reduce the top weight of naval ships, improve ship stability and reduce fuel consumption \cite{Shkolnikov:2014}. Adhesive joints joining dissimilar materials are an integral part of hybrid structure design and have superior properties over mechanical fasteners such as bolts and rivets.  However, the adoption of lightweight materials and adhesive bonding has lagged behind in the shipbuilding industry owing to general concerns regarding their reliability and fatigue lifetime when exposed to harsh marine environments, such as salt water, during their service lifetime \citep{pethrick2015a, kinloch1979a, gettings1977a, bjerk2010a}. A particular concern is the deleterious effect that moisture has upon the strength of a bonded component, especially when under operating or off-normal conditions of high stress and temperature.

At present, the characterization and engineering assessment of adhesive joints relies mainly on laboratory tests \citep{scully1966a, gledhill1974a, schmidt1986a, leung2004a, stratmann1994a, leng1998a, Askarinejad:2022}. The result of these tests is often interpreted and classified, often with remarkable success, using continuum stress-assisted diffusion and fracture-mechanics models \cite{bordes2009a, Askarinejad:2022}. However, the effective material properties required by these models are often not known for new material specifications and need to be calibrated. Testing is costly and time consuming, which impedes exploration and timely deployment of new advanced designs. Therefore, there remains a need to develop a fundamental understanding of the strength degradation mechanisms in adhesive joints in aggressive hygrothermal environments across scales and the means to characterize new material specifications from first principles. 

The aim of the present work is to develop a multiscale model of mechanisms of degradation and failure in adhesive layers enabling qualification of lightweight hybrid structures for safe maritime operation. To this end, we follow and adapt to the present setting the multiscale framework set forth by Serebrinsky, Carter and Ortiz (SCO) \citep{serebrinsky2004a} for predicting hydrogen embrittlement of metals. Whereas the conceptual framework is of quite general applicability, for definiteness in the present work we confine attention to adhesive layers between metallic adherents for which failure is dominated by impurity-dependent cohesive fracture, with impurity uptake from the environment and diffusive transport within the adhesive layer. In this scenario, the SCO framework contemplates the following scales: 

\begin{itemize}
\item[i)] \underline{Microscale:} At the atomic scale, failure happens by irreversible bond breaking, be it in tension or shear or a combination of both. Impurities segregate to the failure planes and attain a concentration per unit area, or coverage, that is in equilibrium with the bulk concentration according to the Langmuir-McLean isotherm \cite{McLean:1957}. Following Rice and Wang \cite{RiceWang:1989}, we regard each failure plane as a thermodynamic system in equilibrium. The attendant Helmholtz free energy per unit area of a failure plane depends on temperature, the sliding displacement and the impurity coverage. By an appeal to linear theory, the sensitivity of the fracture energy to impurity coverage can be expressed in terms of a single property, namely, the difference between the free energy of segregation to a clean bonded atomic plane and the free energy of segregation to a clean free surface \cite{RiceWang:1989}. This property can conveniently be computed from atomistic models using the Born-Haber cycle \citep{ven2003a, jiang2004a}, which opens the way for the characterization of new material specifications from first principles, see Section~\ref{Ow8Ezz} for a discussion. 
\item[ii)] \underline{Mesoscale:} Here, mesoscale refers to the effective behavior of the entire adhesive layer. We consider two main mesoscopic mechanisms: mechanical deformation and degradation of the adhesive layer; and diffusive impurity uptake and transport.
\begin{itemize}
\item[ii.a)] The effective mechanical behavior of the layer is characterized by the collective response of a large number of interatomic planes across the adhesive layer, and can be elucidated by means of a renormalization analysis due to \cite{NguyenOrtiz:2002} that effectively {\sl upscales} atomic-level cohesive behavior to the structural scale (see Appendix~\ref{kRMhEb} for a summary). Alternatively, we show that the same effective behavior of the layer can be derived from a simple energy-release argument. 
\item[ii.b)] For bonded joint configurations, the debonding reaction is mainly limited by the ingress of humidity from the environment and subsequent diffusion to the debonding site \cite{Bonniau:1981}. Adhesives absorb water from the atmosphere through a surface layer that reaches equilibrium quickly with the surrounding environment. The water absorbed consists mainly of molecules or groups of molecules linked by hydrogen bonds to the adhesive (bound water). We assume diffusion to be Fickian (see \cite{borges2021a} for a review of diffusion models). The equilibrium interplanar coverage at the debonding site can then be computed from the bulk concentration in the adhesive using the Langmuir relation. Water diffusion kinetics in epoxy adhesives and in bonded stainless steel/epoxy adhesive interfaces has been characterized, e.g., in \cite{grangeat2022a}, and the effect of temperature and pressure is reported in \cite{duncan2007a}. 
\end{itemize}
\item[iii)] \underline{Macroscale:} The macroscale refers to the structural scale and design configurations thereof. For definiteness, we focus on lap joints under the action of static loads and aggressive hygrothermal environments, (see, e.~g., \cite{leronni2023a}). The goal at this scale is to predict design metrics of interest such as the onset of fracture, the progression of fracture in time and the lifetime of the joint, in terms of material properties and other design parameters. 
\end{itemize}

The resulting multiscale theory predicts that the strength of the adhesive layer depends on thickness, a telltale sign that failure is controlled---at least partially---by fracture. Indeed, the characteristic square-toot scaling of strength as a function of thickness predicted by the theory is found to match the experimental data (e.~g., \cite{Askarinejad:2021b}). 

To make further contact with experiment, as well as facilitating characterization and design, we formulate a simple one-dimensional model of lap joints that can be solved analytically in closed form. The great advantage of analytical tractability with respect to the standard practice of formulating large-scale finite-element models (e.~g., \cite{Korenberg:2004, duncan2007a, Johlitz:2008, bordes2009a, mubashar2009a, bjerk2010a, Diebels:2012, Askarinejad:2021a, Askarinejad:2021b, leronni2022a}) is that it supplies ready-to-use design formulas that elucidate explicitly the dependence of design metrics of interest, such as the lifetime of the joint, on material and structural parameters. 

The simple one-dimensional model predicts that the edge cracks nucleate upon attainment of a threshold that depends on temperature, applied load and environmental chemical potential. The subsequent rate of crack growth is limited by water diffusion kinetics within the adhesive layer and is controlled by the sensitivity of the cohesive strength to water concentration. The resulting length of the edge cracks as a function of time, the lifetime of the joint, and the strength degradation of the joint as a function of time of preexposure to the environment, are all solved in closed form and compared with experiment data where possible. The analytical solution adequately captures the overall trends observed in measurements of moisture concentration in adhesive layers \cite{mubashar2009a}. As expected from the formulation, cracks grow on the diffusive time scale, with characteristic times of the order of days, as often observed experimentally \cite{mubashar2009a, Andreon:2019, Askarinejad:2022}. Remarkably, crack growth rates are controlled by a single variable that combines thermal, mechanical and environmental drivers. The simple one-dimensional model also correctly predicts the gradual degradation attendant to preexposure to conditioning environments, again on the diffusive time scale \cite{kinloch1979a, gledhill1974a, mubashar2009a, bordes2009a}. 

\section{Material modeling}

We are concerned with the lifetime of adhesive joints between metallic plates in contact with an aggressive environment. Diffusion uptake from the environment and subsequent diffusion results in impurity buildup in the cohesive zone of a propagating crack, which decreases the fracture toughness of the material and promotes failure. 

\subsection{Impurity diffusion}

The mechanisms by which impurities are transported to the crack tip are absorption from the environment, bulk diffusion and bond line diffusion. We assume that diffusion in the metal plates is negligible and that the adhesive is strongly bonded to the plates, with the result that mass transport is controlled by bulk diffusion. Under these conditions, the molar bulk chemical potential of the impurities is \cite{PorterEasterling:1981}
\begin{equation} \label{Eq:BulkMu}
  \mu = \mu_0 + R T \ln c - p \, {V} ,
\end{equation}
where $c$ is the impurity mole fraction, $R=8.314$\,J\,/\,mol\,K is the universal gas constant, $T$ is the absolute temperature, $p$ is the pressure, ${V}$ is the partial molar volume of the impurity in solid solution, and $\mu_0$ is the reference chemical potential at $p=0$ and $c=1$. Throughout this work, we take the temperature to be constant. In addition, in the lap test the stress state of the adhesive joint is predominantly shear and we therefore neglect the pressure term in (\ref{Eq:BulkMu}).

The impurity flux ${J}$ is related to the chemical potential by Fick's law, namely, \cite{PorterEasterling:1981}
\begin{equation} \label{Eq:Fick1}
    {J} = - M c \nabla \mu  ,
\end{equation}
where $M$ is the impurity mobility through the adhesive. Inserting (\ref{Eq:BulkMu}) into (\ref{Eq:Fick1}) we obtain
\begin{equation} \label{Eq:Fick1b}
    - 
    {J} 
    =
    D \, \nabla c ,
\end{equation}
where $D=M \, RT$ is the diffusivity of the adhesive. In addition, conservation of mass requires
\begin{equation} \label{Eq:ConservationH}
    - 
    \nabla \cdot {J} 
    = 
    \frac{\partial c}{\partial t} .
\end{equation}
Inserting (\ref{Eq:Fick1b}) into (\ref{Eq:ConservationH}) yields the diffusion equation
\begin{equation} \label{Eq:Fick2}
    \frac{\partial c}{\partial t} = D \, \nabla^2 c .
\end{equation}
which must be satisfied on the bond line.

Impurities enter from the environment at the two end points of the bond line. In order to formulate appropriate boundary conditions at those points, we assume thermodynamic equilibrium between the adhesive and the environment, which requires 
\begin{equation} \label{Eq:FixedBC-mu}
    \mu = \mu_{\rm env}  \ ,
\end{equation}
which, upon insertion of (\ref{Eq:BulkMu}), becomes
\begin{equation} \label{Eq:FixedBC-C}
    c 
    = 
    \exp\Big( \frac{\mu_{\rm env} - \mu_0}{RT} \Big)
    :=
    c_{\rm eq} ,
\end{equation}
where $c_{\rm eq}$ is an equilibrium concentration. 

The equilibrium concentration can in turn be related to the ambient relative humidity $H = e_{\rm env}/e_{\rm sat}(T)$, where $e_{\rm env}$ is the ambient vapor pressure and $e_{\rm sat}(T)$ is the saturation vapor pressure at temperature $T$. Using the equilibrium relation \cite{Campbell:1998}
\begin{equation}
    \frac{\mu_{\rm env} - \mu_0}{RT}
    =
    \log\frac{e_{\rm env}}{e_{\rm sat}(T)}
    =
    \log H
\end{equation}
and (\ref{Eq:FixedBC-C}), it follows that $c_{\rm eq} \sim H$, i.~e., the equilibrium concentration is measured by the relative humidity, which is the environmental metric often reported in experimental studies (see, e.~g., Section~\ref{wWCcio}). 

Standards for the measurement and characterization of absorption and diffusion of moisture in polymeric materials, which include plastics, polymer matrix composites and adhesives, are discussed in detailed in \cite{duncan2007a}. Whereas the Fickian model just described is most commonly used in practice, a number of other models, such as the Langmuir model, the dual Fickian model and others are also utilized when the simple Fickian model proves inadequate (see \cite{borges2021a} for a review).

\subsection{Cohesive laws accounting for impurity segregation} \label{ioqs4Z}

Cohesive theories of fracture, pioneered by Dugdale \cite{Dugdale:1960}, Barrenblatt \cite{Barenblatt:1962}, Rice \cite{Rice:1968} and others, regard fracture as a gradual process in which separation and sliding take place across an extended crack tip, or cohesive zone, and is resisted by tractions according to a cohesive law. Of particular concern in the present setting is the deleterious effect of impurities on the cohesive law. In this work, we account for that effect through a appeal to the thermodynamic theory of interface embrittlement by solute segregation of Rice and Wang \cite{RiceWang:1989}. 

\subsubsection{Atomic-level decohesion}

At the atomic level, decohesion happens by irreversible bond breaking, be it in tension or shear or a combination of both. In particular, we shall assume that bonds broken in shear do not heal upon further interfacial sliding. We additionally assume that the adhesive layer slides by $\delta$ under the action of an applied shear stress $\tau$ along a large number of possible atomic failure planes parallel to the layer; and that the impurities segregate to the failure planes and attain a concentration $\Gamma$ per unit area that is in equilibrium with the bulk concentration $c$. Following \cite{RiceWang:1989}, we regard each failure plane as a thermodynamic system in equilibrium. The Helmholtz free energy $f$ of a failure plane per unit area then obeys the equilibrium relation 
\begin{equation} \label{eq:df}
    df = \tau d\delta + s dT + \mu d\Gamma ,
\end{equation}
where $T$ is the absolution temperature, $s$ is the entropy per unit area of failure plane and $\mu$ is the chemical potential. The key assumption underlying this relation is that all thermodynamic functions, such as $f$, $\tau$, $s$ and $\mu$ depend on the variables $(\delta;T,\Gamma)$. 

An important parameter controlling decohesion is the ideal work $2 \gamma$ of irreversible bond breaking per unit area. Under isothermal conditions and fixed composition, it follows from (\ref{eq:df}) that
\begin{equation} \label{wMsTwR}
    2 \gamma(T,\Gamma)
    = 
    f(+\infty;T,\Gamma) - f(0;T,\Gamma)
    =
    \int_0^{+\infty} \tau(\delta;T,\Gamma) \, d\delta ,
\end{equation}
i.~e., $2\gamma$ equals the area under the cohesive law governing the relation between the stress a vs. displacement-jump across the failure plane. We note that $f(+\infty;T,\Gamma)$ = $2 f_s(T,\Gamma/2)$, where $f_s(T,\Gamma/2)$ is the free energy per unit area of a free surface under the assumption that each of the two free surfaces generated by a failure plane receive one half of the impurities in the plane. In addition $f(0;T,\Gamma)$ is the free energy per unit area of one unstressed failure plane. 

\subsubsection{Renormalization and upscaling of atomic-scale cohesive laws}

It is not possible to utilize atomic-level cohesive laws directly at the macroscale. Thus, atomic-scale interplanar potentials are characterized by peak stresses of the order of the theoretical strength of the material. In addition, the materials lose their bearing capacity after an interplanar separation of only a few angstroms. The cohesive-zone sizes attendant to first-principles interplanar potentials are also on the nanometer scale, which renders them unusable at the macroscale. 

This disconnect is resolved by recognizing that the effective macroscopic behavior of an adhesive layer entails the equilibrium response of a large number of possible atomic-level failure planes. Nguyen and Ortiz \cite{NguyenOrtiz:2002} put forth a renormalization procedure that characterizes the macroscopic equilibrium response of a macroscopic system and effectively  upscales atomic-level cohesive behavior to the macroscale. A summary account of the analysis is included in Appendix~\ref{kRMhEb} for completeness. 

For a layer of thickness $h \gg \ell$, with $\ell$ a typical atomic separation between failure planes, it follows from the analysis that, asymptotically, the macroscopic system is governed by the simple universal law
\begin{equation} \label{YEpmwv}
    {\tau}({\delta}) =
    \begin{cases}
    {\tau}_c \, {\delta}/{\delta}_c ,
    & \mbox{if ${\delta} < {\delta}_c$}  \\
    0,  & \mbox{otherwise} ,
    \end{cases}
\end{equation}
where ${\tau}$ is the macroscopic shear stress on the layer, ${\delta}$ is the total sliding displacement of the layer, and
\begin{equation} \label{sCfg6h}
    {\tau}_c = 2 \sqrt{\frac{{G} \, \gamma}{h}} ,
    \qquad
    {\delta}_c = 2 \sqrt{\frac{\gamma \, h}{{G}}} ,
\end{equation}
are the renormalized shear strength and critical sliding displacement of the layer, respectively, which follow explicitly in terms of the atomic-level specific fracture energy per unit area $2\gamma$, the elastic shear modulus ${G}$ and, crucially, the thickness $h$ of the layer. The corresponding macroscopic free energy per unit area is
\begin{equation} \label{l7uB4f}
    {f}_0({\delta}) =
    \begin{cases}
    2 \gamma \, ( {\delta}/{\delta}_c )^2 ,
    & \mbox{if ${\delta} < {\delta}_c$}  \\
    2 \gamma,  & \mbox{otherwise} .
    \end{cases}
\end{equation}
Thus, the renormalized traction-separation law is elastic up to a critical shear stress ${\tau}_c$, and subsequently drops to zero upon the attainment of a critical sliding displacement ${\delta}_c$. This renormalization procedure has been verified numerically \cite{HayesOrtizCarter:2004} and analyzed mathematically \cite{BraidesLewOrtiz:2006}. Numerical experiments \cite{HayesOrtizCarter:2004} based on {\sl ab initio} calculations show that an atomic layer of thickness $h \gtrsim 10 \ell$ exhibits cohesive behavior ostensibly indistinguishable from the renormalized law (\ref{YEpmwv}). 

The relations (\ref{sCfg6h}) can also be derived from a simple energy-release argument. Thus, assume a single shear crack growing on the middle plane of the adhesive layer deforming under the action of a uniform sliding displacement $\delta$ on its boundary. Far ahead of the crack tip the energy density of adhesive layer is
\begin{equation}
    {W}_c = \frac{G}{2}\Big( \frac{{\delta}_c}{h} \Big)^2 ,
\end{equation}
whereas far behind the crack tip the energy density is ${W} = 0$. Therefore, the energy-release rate per unit crack advance is $d {W}_c$, whence an appeal to Griffith's criterion gives the relation
\begin{equation}
    \frac{G h}{2} \, \Big( \frac{{\delta}_c}{h} \Big)^2 = 2 \gamma ,
\end{equation}
and the second of (\ref{sCfg6h}) is recovered. The first of (\ref{sCfg6h}) then follows from the relation ${\tau}_c = G {\delta}_c/h$. 

A remarkable outcome of the renormalization analysis is that the asymptotic form (\ref{YEpmwv}) and (\ref{l7uB4f}) of the effective macroscopic cohesive law and cohesive free energy per unit area, respectively, are {\sl universal} and depends only on the elastic modulus ${G}$, the specific fracture energy $2\gamma$ of one atomic-level failure plane and the thickness $h$ of the layer, which greatly facilitates characterization, e.~g., by recourse to atomistic calculations (refer to Section~\ref{Ow8Ezz} for discussion and outlook). 

\subsection{Static failure of MMA adhesive layer}

\begin{table}
\centering
%\begin{center}
\begin{tabular}{|c|c|c|c|c|c|}
  \hline
  $h$ (mm) &
  3 & 
  5 & 
  8 & 
  10 & 
  13 \\ \hline
  $\tau_c$ (MPa) &
  12.34 & 
  11.95 & 
  11.41 & 
  10.61 & 
  8.75 \\  \hline
\end{tabular}
\caption{Shear strength vs. layer thickness data extracted from Fig.~4a of \cite{Askarinejad:2021b}.}\label{iKNQmO}
%\end{center}
\end{table}

By way of validation of the cohesive strength model just outlined, we compare the thickness-scaling laws (\ref{sCfg6h}) to data from \cite{Askarinejad:2021b} concerned with the adhesive fracture of Thick-Adherend-Shear-Test (TAST) specimens made from a MMA-based adhesive (Scigrip SG300-40) and steel adherents. The data show that the value of the shear stress corresponding to the onset of crack growth, which we assume to scale with the cohesive strength $\tau_c$ of the adhesive layer, decreases with increasing layer thickness, see \cite[Fig.~4(a)]{Askarinejad:2021b}. Corresponding values of thickness and strength gleaned from that figure are tabulated in Table~\ref{iKNQmO}. 

\begin{figure}[ht!]
\begin{center}
    \includegraphics[width=0.69\textwidth]{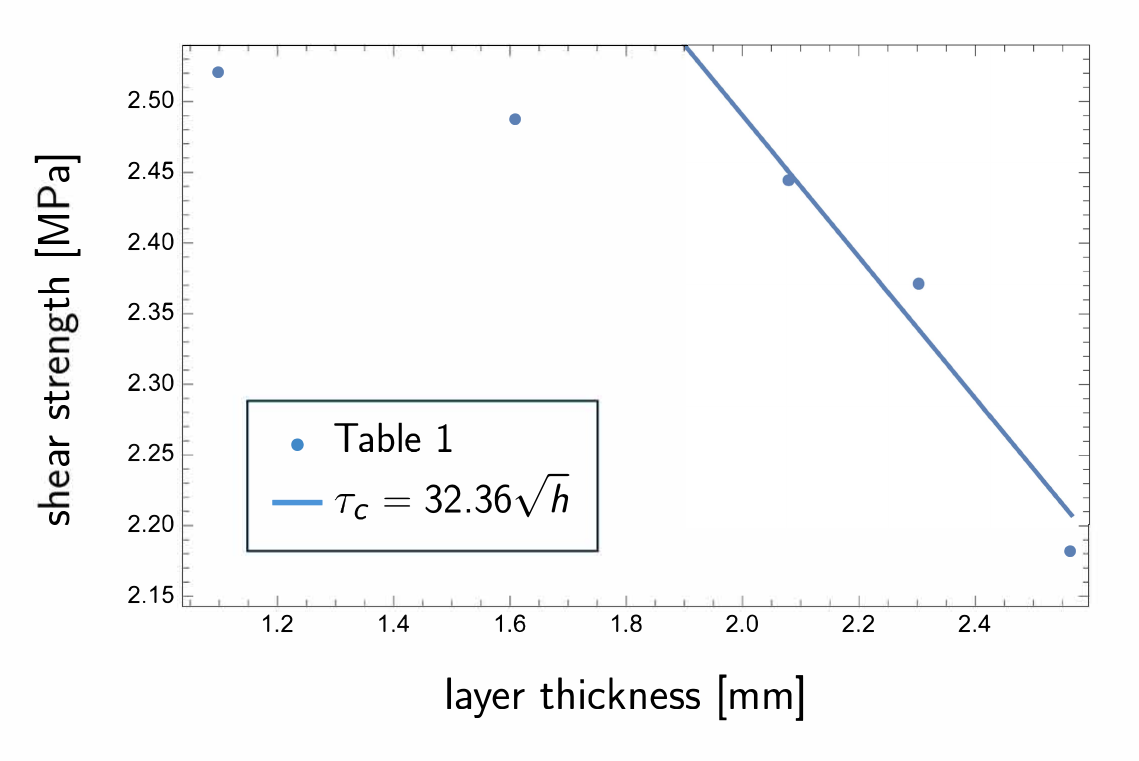}
    \caption{Log-log fit of strength {\sl vs}.~thickness data in Table~\ref{iKNQmO} (extracted from Fig.~4a of \cite{Askarinejad:2021b}).} \label{ZfT3Ez}
\end{center}
\end{figure}

Fig.~\ref{ZfT3Ez} shows the data and a fit of the first of (\ref{sCfg6h}) in the large thickness range, namely,
\begin{equation} \label{zGq6Yi}
    \tau_c[{\rm MPa}] 
    = 
    32.36[{\rm MPa} \sqrt{\rm mm}]/\sqrt{h[{\rm mm}]} .
\end{equation}
As may be seen from the comparison, the scaling laws (\ref{sCfg6h}) adequately capture the asymptotic trend of the data for large thicknesses, as expected. Assuming a Young's modulus $E \sim 300$ MPa \cite{Askarinejad:2022} and a typical Poisson's ratio $\nu \sim 0.38$ for MMA \cite{Moller:2015}, or $G \sim 100$ MPa, and matching constants between (\ref{sCfg6h}) and (\ref{zGq6Yi}), we obtain $2\gamma = 5.38$ MPa/m${}^2$, which is within the ballpark of MMA fracture energies \cite{Moller:2015}. 

\subsection{Impurity-dependent cohesive behavior}

We recall that thermodynamic arguments give $2\gamma(T,\Gamma)$ as a function of the temperature $T$ and the impurity concentration $\Gamma$ per unit area of the failure plane, eq.~(\ref{wMsTwR}). For instance, assuming linear response,
\begin{equation} \label{igM3aT}
    2\gamma(T,\Gamma) \sim 2\gamma_0 - \Delta g_0 \Gamma ,
\end{equation}
where $2\gamma_0$ is the work to fail a clean atomic plane, and $\Delta g_0 = \Delta g_{0,b} - \Delta g_{0,s}$ is the difference between the free energy of segregation $\Delta g_{0,b}$ to a clean bonded atomic plane and the free energy of segregation $\Delta g_{0,s}$ to a clean free surface \cite{RiceWang:1989}. For normal adhesives, we expect $\Delta g_0 > 0$, or $\Delta g_{0,b} > \Delta g_{0,s}$, i.~e., that it is energetically favorable to segregate impurities to a clear free surface than to a clean bonded atomic plane. Under these conditions, impurities have a deleterious effect on the specific fracture energy of the adhesive, as expected. Inserting the function $2\gamma(T,\Gamma)$ into (\ref{l7uB4f}) gives an impurity-dependent macroscopic cohesive free-energy per unit area ${f}_0({\delta}; T, \Gamma)$, which depends parametrically on $(T, \Gamma)$. 

The requisite closure relation between $\Gamma$ and the macroscopic impurity concentration ${c}$ then follows by an appeal to the Langmuir-McLean isotherm \cite{McLean:1957}
\begin{equation} \label{Kbvdxf} 
    \Gamma = \Gamma_{\rm max}
    \frac{c_{\rm bulk} V}{c_{\rm bulk} V + \lambda} ,
    \quad
    \lambda
    :=
    \exp(-\Delta h/RT) ,
\end{equation}
where $c_{\rm bulk}$ is the impurity concentration in the bulk, ${V}$ is the partial molar volume of the impurity in solid solution, $\Gamma_{\rm max}$ is the saturation value of $\Gamma$, $\Delta h$ is the enthalpy difference between the bulk and the adsorbed states. In addition, if, as before, $\ell$ is the separation between possible failure planes and $h$ is the thickness of the layer, we have the relation
\begin{equation} 
    {c} = \frac{1}{\ell} \Gamma + c_{\rm bulk} .
\end{equation}
Eliminating $c_{\rm bulk}$ between these two relations gives
\begin{equation} \label{5n0TXQ} 
    \Gamma = 
    \frac{-\sqrt{(\ell (\lambda -{c} V)+\Gamma_{\rm max} V)^2+4
    {c} \ell^2 \lambda  V}+{c} \ell V+\Gamma_{\rm max} V+\ell
    \lambda }{2 V} ,
\end{equation}
which allows the impurity-dependent macroscopic cohesive free-energy per unit area ${f}_0({\delta}; T, \Gamma)$ to be expressed as a function ${f}_0({\delta}; T, {c})$
of the macroscopic impurity concentration. 

For dilute concentrations, to first order (\ref{5n0TXQ}) reduces to
\begin{equation}
    \Gamma \sim
    \frac{\Gamma_{\rm max} V / \ell}{\Gamma_{\rm max} V / \ell + \lambda } 
    \, \ell  \, {c} .
\end{equation}
Finally, if the impurities bond strongly to the failure planes, $\Delta g/RT \gg 1$, then $c_{\rm bulk} \sim 0$, i.~e., the bulk depletes of impurities, and 
\begin{equation} \label{XP5Fcm}
    \Gamma \sim \ell  \, {c} ,
\end{equation}
corresponding to all the impurities segregating to the failure planes. 

The combination of (\ref{XP5Fcm}) and (\ref{igM3aT}) suggests, to a first approximation, a linear dependence of the specific fracture energy of the adhesive on impurity concentration. Indeed, this type of relation is born out by experiment. For instance, Korenberg {\sl et al.} \cite{Korenberg:2004} measured a linear relationship between specific fracture energy, and relative humidity (RH) for the grit-blast and degrease (GBD) pretreated aluminium alloy and steel joints bonded with a rubber-toughened epoxy adhesive.

\subsection{Application to the BADGE/DDS adhesive} \label{wWCcio}

As an example of application of the theory, we estimate the fundamental properties $2\gamma_0$ and $\Delta g_0$ in (\ref{igM3aT}) for the BADGE/DDS MMA system \cite{Prolongo:2006, Dyer:2024}. BADGE (Bisphenol A Diglycidyl Ether) is the basic monomer unit of an epoxy resin. It is the main building block used to create the final epoxy-based polymer. In its pure form, BADGE is a viscous, colorless to pale yellow liquid. It's widely used in various industrial applications like coatings, adhesives, and sealants due to its excellent properties. DDS (4-Aminophenyl sulfone) is a hardener or curing agent that is used with epoxy resins like BADGE. Hardeners are mixed with the epoxy resin (BADGE) in a specific ratio to initiate the "curing" or "hardening" process, which forms the final, solid epoxy-based polymer. 

\begin{table} 
\centering
%\begin{center}
\begin{tabular}{|c|c|c|}
  \hline
  $T$ (${}^\circ$C) & RH (\%) & ${\tau}_c$ (MPa) \\ \hline\hline
  23 & 30 & 5.8 $\pm$ 0.3 \\ \hline
  23 & 80 & 5.6 $\pm$ 0.3 \\ \hline
  80 & 30 & 4.8 $\pm$ 0.2 \\ \hline
  80 & 80 & 4.4 $\pm$ 0.5 \\ \hline
\end{tabular}
\caption{Adhesive strength of BADGE/DDS adhesive at different temperatures and relative humidities \cite[Table 2]{Prolongo:2006}.} \label{ZlXpQJ}
%\end{center}
\end{table}

\begin{table}
\centering
%\begin{center}
\begin{tabular}{|c|c|c|c|c|c|}
  \hline
  Tensile &
  Tensile & 
  Flexural & 
  Fracture & 
  Free & 
  Density \\
  modulus &
  strength & 
  modulus & 
  toughness & 
  volume & 
  (kg/cm${}^3$) \\
  (GPa) &
  (MPa) & 
  (GPa) & 
  (MPa$\sqrt{\rm m}$) & 
  (\AA) & 
  ${}$ \\ \hline\hline
  1.30 & 72.09 & 2.50 & 0.51 & 2.62 & 1217 \\ \hline
\end{tabular}
\caption{Mechanical properties of BADGE/DDS adhesive system at room temperature \cite[Table 1]{Dyer:2024}.} \label{rEBw25}
%\end{center}
\end{table}

Typical experimental data for the BADGE/DDS system is reported in \cite{Prolongo:2006, Dyer:2024} and collected in Tables~\ref{ZlXpQJ} and \ref{rEBw25}. From the first of (\ref{sCfg6h}), 
\begin{equation} \label{No1P5O}
    2 \, \gamma = \frac{{\tau}_c^2 h}{2 G} .
\end{equation}

The shear modulus is $G = E/2(1+\nu)$. From Table~\ref{rEBw25}, $E=1.3$ GPa. Taking $\nu=0.4$ as a typical value of the Poisson's ratio \cite{Pandini:2008, Reedy:1997}, we obtain $G = 0.5$ GPa. In addition, $h = 0.22 \pm 0.02$ mm \cite{Prolongo:2006}. These properties and (\ref{sCfg6h}) allow $\tau_c$ to be replaced by corresponding values of $2 \gamma$  in Table~\ref{ZlXpQJ}. % $2 \gamma = 2 \times 10^{-5} {\tau}_c^2$.
Further inserting (\ref{No1P5O}) into (\ref{igM3aT}) yields the relation
\begin{equation}
    2\gamma_0 - \Delta g_0 \Gamma = \frac{{\tau}_c^2 h}{2 G} ,
\end{equation}
which, in the linear regime under consideration, gives $2\gamma_0$ and $\Delta g_0$ from two corresponding values of $\Gamma$ and ${\tau}_c$. Further identifying $c_{\rm eq} \sim RH/100$ (see Section~\ref{ioqs4Z}) and taking $\ell \sim 2.62$ \AA, Table~\ref{rEBw25}, eq.~(\ref{XP5Fcm}) gives estimates of the coverage $\Gamma$. 

\begin{table} 
\centering
%\begin{center}
\begin{tabular}{|c|c|c|}
  \hline
  $T$ (${}^\circ$C) & $2\gamma_0$ (Jm${}^{-2}$) & $\Delta g_0$ (GPa) \\ \hline\hline
  23 & 7.70 & 3.83 \\ \hline
  80 & 5.55 & 6.18 \\ \hline
\end{tabular}
\caption{Values of $2\gamma_0$ and $\Delta g_0$ in (\ref{igM3aT}) estimated from the data of Tables~\ref{ZlXpQJ} and \ref{rEBw25}.} \label{Huzyrs}
%\end{center}
\end{table}

The values of $2\gamma_0$ and $\Delta g_0$ estimated through these relations from the data of Tables~\ref{ZlXpQJ} and \ref{rEBw25} are tabulated in Table~\ref{Huzyrs}, which showcases the feasibility of calibrating the impurity-dependent cohesive model from standard experimental data. As expected, the specific fracture energy decreases with increasing temperature and with increasing humidity. However, we note that the data are sparse and may not be available for new material specifications, which suggests the use of first-principles or molecular dynamics calculations for purposes of material characterization. 

\section{Application to lifetime prediction of single lap-joint specimens}

\begin{figure}[ht!]
\begin{center}
    \includegraphics[width=0.66\textwidth]{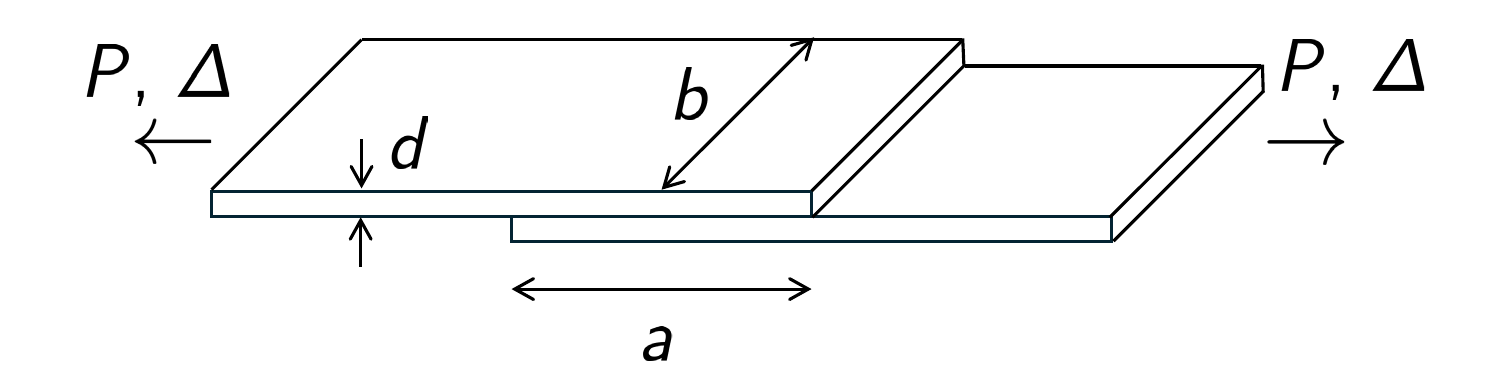}
    \caption{Lap test geometry considered in the lifetime analysis.} \label{FIDpWS}
\end{center}
\end{figure}

We apply the preceding model to lifetime prediction in a lap-joint test configuration, Fig.~\ref{FIDpWS}. The analysis operates entirely at the macroscopic level. In particular, the behavior of the adhesive layer is assumed to be fully characterized by the impurity-dependent effective free energy density per unit area (\ref{l7uB4f}) and (\ref{sCfg6h}), with impurity-dependent specific fracture energy in the linear response range (\ref{igM3aT}) and closure relation (\ref{XP5Fcm}). 
In addition, we restrict attention to isothermal conditions and omit any and all dependencies on temperature for simplicity of notation.

\subsection{Impurity profile}

Since the adhesive layer in the lap test deforms predominantly in shear, and no normal opening displacement as might allow for infiltration of the bond line occurs, we shall assume that the mobility $M$ in (\ref{Eq:Fick1}) and equilibrium condition (\ref{Eq:FixedBC-C}) at the bond-line end points remain independent of the sliding displacement along the bond line. By virtue of this simplification, the diffusion problem becomes decoupled from deformation and can be solved independently of the mechanics problem.

\begin{figure}[ht!]
\begin{center}
    \includegraphics[width=0.82\textwidth]{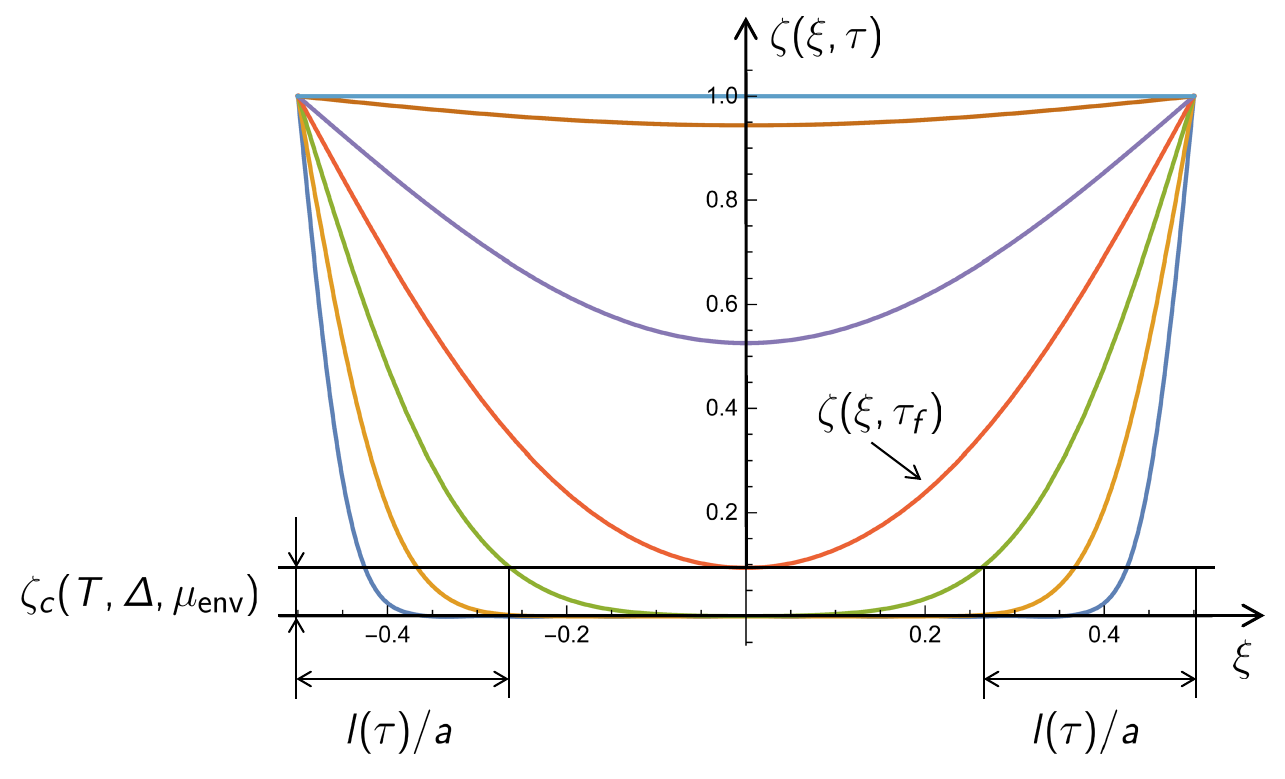}
    \caption{Time evolution of impurity profile over bond line and construction for determining the evolution of the crack length $l(t)$. Normalized concentration profiles $\zeta(\xi,\tau)$ shown at normalized times $\tau = 0.001$, $0.001\sqrt{10}$, $0.01$, $0.01\sqrt{10}$, $0.1$, $0.1\sqrt{10}$ and $1$. At any given normalized time $\tau$, edge cracks extend to the point where the normalized concentration $\zeta(l(\tau)/a,\tau)$ attains a critical value $\zeta_c(T,\Delta,\mu_{\rm env})$ that combines the effects of temperature $T$, applied displacement $\Delta$ and environmental chemical potential $\mu_{\rm env}$. The edge cracks span the entire length of the bond line at the normalized time $\tau_f$, signaling the failure of the joint and the end of its lifetime. } \label{VSl8dk}
\end{center}
\end{figure}

Assuming one-dimensional diffusion over the bond line, the governing initial-boundary-value problem is
\begin{subequations} \label{t6eysK}
\begin{align}
    &   \label{t6eysK1}
    c_t(x,t) = D \, c_{xx}(x,t), 
    \quad
    -a/2 \leq x \leq a/2,
    \quad
    t > 0 ,
    \\ &
    - 
    c(-a/2,t) = c_{\rm eq} ,
    \quad
    c(a/2,t) = c_{\rm eq} ,
    \\ &
    c(x,0) = 0, 
    \quad
    -a/2 \leq x \leq a/2,
\end{align}
\end{subequations}
where $x$ is measured from the center of the bond line, $a$ is the length of the lap joint, and $c(x,t)$ is the impurity concentration at position $x$ and time $t$. Upon normalization, 
\begin{equation} \label{o8lX7F}
    \xi := \frac{x}{a} ,
    \quad
    \tau := \frac{D t}{a^2} ,
    \quad
    \zeta := \frac{c}{c_{\rm eq}} ,
\end{equation}
problem (\ref{t6eysK}) becomes
\begin{subequations} 
\begin{align}
    &   \label{pOqTNh}
    \zeta_\tau(\xi,\tau) = \zeta_{\xi\xi}(\xi,\tau), 
    \quad
    -1/2 \leq \xi \leq 1/2,
    \quad
    \tau > 0 ,
    \\ &
    \zeta(-1/2,\tau) = 1 ,
    \quad
    \zeta(1/2,\tau) = 1 ,
    \\ &
    \zeta(\xi,0) = 0, 
    \quad
    -1/2 \leq \xi \leq 1/2 .
\end{align}
\end{subequations}
In normalized quantities the solution is \cite{Crank:1975}
\begin{equation} \label{0vg5Kc}
    \zeta(\xi,\tau)
    =
    1
    -
    \frac{2}{\pi} \, 
    \sum_{k=1}^\infty
    \frac{1+(-1)^{k+1}}{k} \,
    {\rm e}^{-\pi ^2 k^2 \tau} \,
    \sin(\pi k (\xi+\nicefrac{1}{2})) ,
\end{equation}
which fully characterizes the transient impurity profile over the bond line, Fig~\ref{VSl8dk}. In particular, at the center of the joint overlap, $\xi = 0$, the sum evaluates explicitly to
\begin{equation} \label{PHI89N}
    \zeta(0,\tau)
    =
    1
    -
    {\rm erf}
    \Big(
        \frac{1}{2\sqrt{\tau}}
    \Big) 
    =
    {\rm erfc}
    \Big(
        \frac{1}{2\sqrt{\tau}}
    \Big) ,
\end{equation}
where ${\rm erf}(x)$ and ${\rm erfc}(x)$ are the error function and the complementary error function, respectively \cite{Abramowitz:1964}. The analytical solution (\ref{0vg5Kc}) adequately captures the overall trends observed in measurements of moisture concentration in adhesive layers \cite{mubashar2009a}.

\subsection{Crack growth}

Suppose, for simplicity, that the plates in the joint are much stiffer than the adhesive. Then, to a first approximation we can take the plates to be rigid, whereupon the sliding displacement follows simply as
\begin{equation}
    \delta(x,t) = 2 \Delta = \text{const} ,
\end{equation}
and the position of the crack tips is determined by the condition
\begin{equation}
    2 \Delta = \delta_c(\pm a/2-l(t),t) . 
\end{equation}
Inserting the second of (\ref{sCfg6h}) into this condition, we obtain
\begin{equation}
    \Delta 
    =
    \sqrt{\frac{\gamma(c(a/2-l(t),t)) \, h}{{G}}} ,
\end{equation}
where $c(x,t)$ is given by the solution (\ref{0vg5Kc}) in terms of normalized quantities (\ref{o8lX7F}) and the dependence of $2\gamma$ on the local impurity concentration $c(x,t)$ follows from (\ref{igM3aT}), (\ref{Kbvdxf}) and (\ref{5n0TXQ}). Using the relations (\ref{igM3aT}) and (\ref{XP5Fcm}), we obtain
\begin{equation}
    \Delta 
    =
    \sqrt
    {
        \frac
        { 
            (\gamma_0 - \Delta g_0  c(a/2-l(t),t) \ell/2) \, h
        }
        {
            {G}
        }
    } ,
\end{equation}
whence the concentration at the crack tip follows as
\begin{equation}
    c(a/2-l(t),t)
    =
    \frac{2}{\ell}
    \frac{\gamma_0 - (G/h) \Delta^2 }{\Delta g_0}
    :=
    c_c(T,\Delta,\mu_{\rm env}) ,
\end{equation}
or, in normalized quantities,
\begin{equation} \label{OTaTZT}
    \zeta(\nicefrac{1}{2}-l(t)/a,\tau)
    =
    \frac{2}{\ell}
    \frac{\gamma_0 - (G/h) \Delta^2 }{c_{\rm eq} \Delta g_0}
    :=
    \zeta_c(T,\Delta,\mu_{\rm env})
\end{equation}
which can be solved for $l(t)/a$. 

\begin{figure}[ht!]
\begin{center}
    \includegraphics[width=0.69\textwidth]{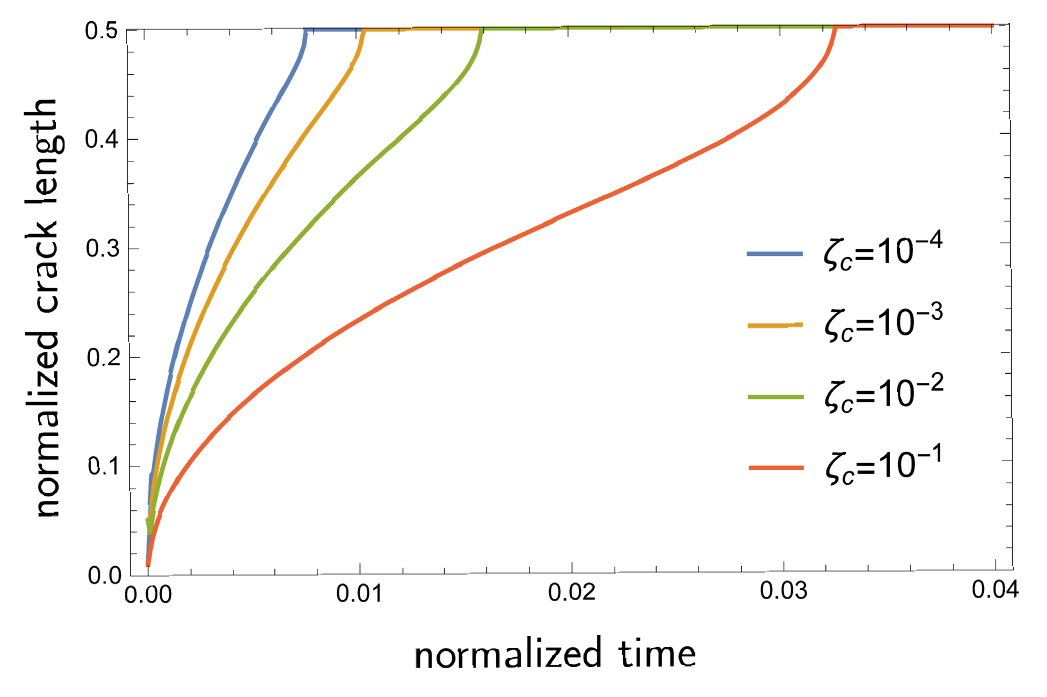}
    \caption{Normalized crack length {\sl vs}.~normalized time for various values of the effective driving force $1-\zeta_c(T,\Delta,\mu_{\rm env})$.} \label{p9ZBhN}
\end{center}
\end{figure}

A graphical representation of (\ref{OTaTZT}), depicting the crack length as a function of time, loading and environmental conditions, is shown Fig.~\ref{p9ZBhN}. Owing to the steady degradation of the adhesive attendant to the uptake of impurities, the critical sliding displacement $\delta_c(x,t)$ expressed as a function of position decreases in time, allowing the edge cracks to progress inward. The resulting time dependence of the crack length exhibits three well-differentiated stages: I) A first stage of rapid growth; II) an intermediate stage of steady linear growth; and III) a final stage of accelerated growth. For short times, the relation (\ref{OTaTZT}) reduces asymptotically to
\begin{equation} \label{jTXbG1}
    1
    -
    \frac{l(t)/a}{\sqrt{\pi\tau}} 
    \sim
    \zeta_c(T,\Delta,\mu_{\rm env}) ,
\end{equation}
or, undoing normalization, to
\begin{equation} \label{EFOSnV}
    l(t)
    \sim
    (1-\zeta_c(T,\Delta,\mu_{\rm env})) \,
    \sqrt{\pi D t} .
\end{equation}
Thus, at early times the crack length $l(t)$ is predicted to grow as $\sqrt{t}$, Fig.~\ref{p9ZBhN}, which is born out by experiment \cite{Leng:1999, Andreon:2019, Askarinejad:2022}. 

It is remarkable that crack growth rates are controlled by a single effective parameter $\zeta_c(T,\Delta,\mu_{\rm env})$ that combines thermal, mechanical and environmental drivers. In addition, under the assumed conditions of impurity-dependent cohesive fracture and Fickian diffusion of the impurities, it follows that cracks grow on the diffusive time scale $\sim a^2/D$. For instance, for joint sizes in the range $a = 1$ to $10$ cm and a diffusion coefficient $D \sim 10^{-12}$ m${}^2$/s, the characteristic time for the evolution of the moisture content is the range $\sim 10^8$--$10^{10}$ s $\sim 1.16$--$116$ days, which is in the ballpark of experimental observation \cite{mubashar2009a, Andreon:2019, Askarinejad:2022}.

\subsection{Lifetime prediction}

\begin{figure}[ht!]
\begin{center}
    \includegraphics[width=0.69\textwidth]{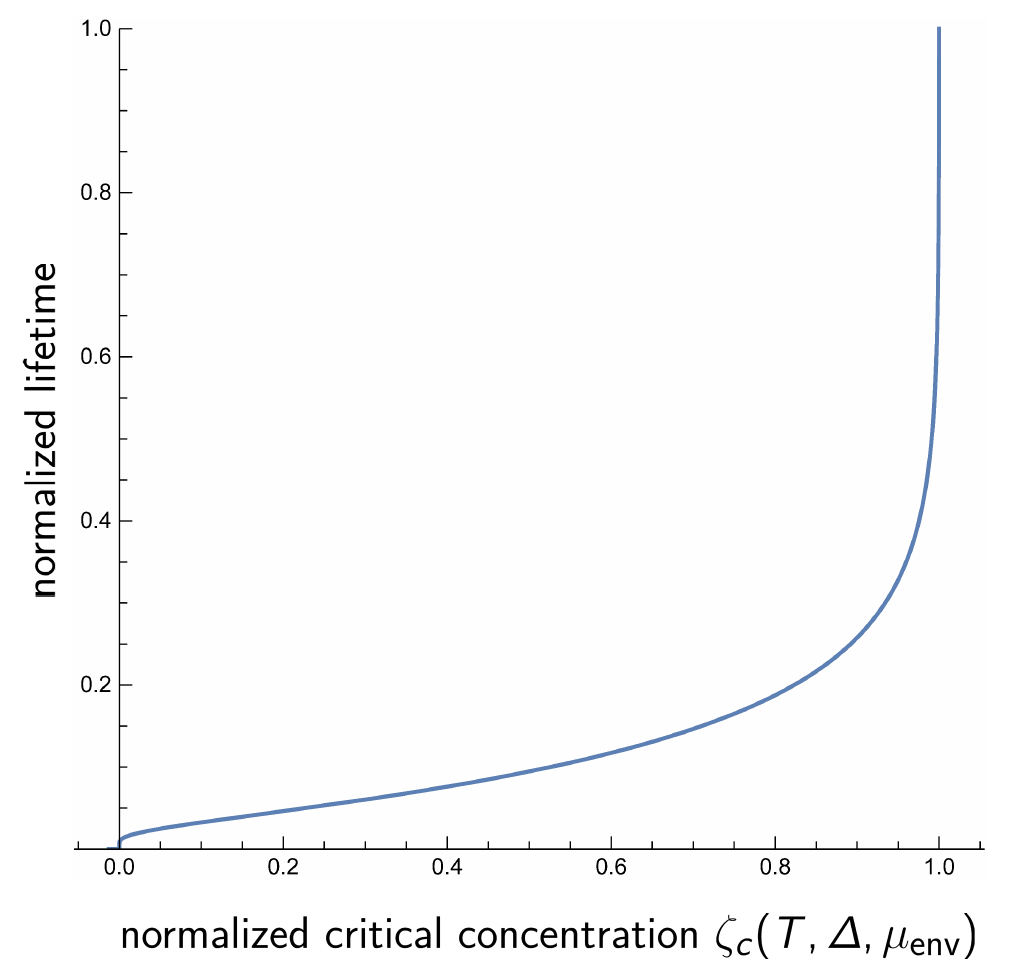}
    \caption{Normalized lifetime of the lap joint as a function of the normalized critical concentration $\zeta_c(T,\Delta,\mu_{\rm env})$.} \label{i4s2fp}
\end{center}
\end{figure}

The joint fails at a time $t_f$, the lifetime of the joint, such that
\begin{equation}
    l(t_f) = a/2 ,
\end{equation}
i.~e., when the edge cracks grow to span the entire joint. Inserting (\ref{PHI89N}) into (\ref{OTaTZT}) with $l(t_f) = a/2$ and solving for $t_f$ undoing normalization (\ref{o8lX7F}) gives, explicitly,
\begin{equation} \label{vgi36M}
    t_f
    =
    \frac{a^2}{D}
    \left(
        \frac{1}{2}
        \frac
        {
            1
        }
        {
            {\rm erf}^{-1}
            \big(
                1 
                -
                \zeta_c(T,\Delta,\mu_{\rm env})
            \big)
        }
    \right)^2
\end{equation}
where ${\rm erf}^{-1}(x)$ denotes the inverse error function. An asymptotic expansion for large $t_f$, or $\zeta_c(T,\Delta,\mu_{\rm env}) \sim 1$, gives 
\begin{equation}
    1-\frac{2}{\pi} {\rm e}^{- \pi^2 \tau_f}
    \sim
    \zeta_c(T,\Delta,\mu_{\rm env}) ,
\end{equation}
or, solving and undoing normalization,
\begin{equation} \label{Cv7WOD}
    t_f
    \sim
    \frac{a^2}{\pi^2 D}
    \log\Big(\frac{\pi}{2} \frac{1}{1-\zeta_c(T,\Delta,\mu_{\rm env})} \Big) .
\end{equation}

The normalized lifetime of the adhesive joint {\sl vs}.~$\zeta_c(T,\Delta,\mu_{\rm env})$ is shown in Fig.~\ref{i4s2fp}. Again, it is remarkable that crack growth rates are controlled by a single effective driving force $1-\zeta_c(T,\Delta,\mu_{\rm env})$ that combines thermal, mechanical and environmental drivers. In particular, for small values of $1-\zeta_c(T,\Delta,\mu_{\rm env})$, or values of $\zeta_c(T,\Delta,\mu_{\rm env})$ close to $1$, the lifetime of the joint is long, whereas for large values of $1-\zeta_c(T,\Delta,\mu_{\rm env})$, or values of $\zeta_c(T,\Delta,\mu_{\rm env})$ close to $0$, the lifetime of the joint is comparatively short. 

The analysis also identifies three distinct regimes:
\begin{itemize}
\item[i)] {\sl Instantaneous failure} occurs if
\begin{equation} \label{kEHjSl}
    \frac{2G}{h} \Delta^2 \geq 2\gamma_0 ,
\end{equation} 
corresponding to the attainment of Griffith's criterion for static fracture. 

\item[ii)] {\sl Delayed crack growth} occurs if
\begin{equation}
    \frac{2G}{h} \Delta^2 < 2\gamma_0 
    \quad
    \text{and}
    \quad
    \ell c_{\rm eq} \Delta g_0
    + 
    \frac{2G}{h}  \Delta^2
    >
    2\gamma_0 .
\end{equation}

\item[iii)] {\sl No crack growth}, instantaneous or delayed, occurs if 
\begin{equation} \label{YQ1c5Y} 
    \ell c_{\rm eq} \Delta g_0
    +
    \frac{2G}{h} \Delta^2 
    \leq 
    2 \gamma_0 .
\end{equation}
\end{itemize}

We see from these alternatives that fracture, and eventually failure, requires a particular combination of environmental factors, represented by the environmental chemical potential $\mu_{\rm env}$, temperature and loading, represented by the prescribed displacement $\Delta$, to exceed a material-specific threshold. 

\subsection{Strength degradation}

The threshold conditions (\ref{kEHjSl}) and (\ref{YQ1c5Y}) set forth two measures of strength that are useful for design and for making contact with test data, which is often reduced to measures of strength. Thus, from (\ref{kEHjSl}),
\begin{equation} 
    \Delta_0 = \sqrt{\frac{\gamma_0 h}{G}} ,
\end{equation} 
arises as a critical displacement for instantaneous failure of the joint. Using the relation $\tau \sim 2G \Delta/h$, we find 
\begin{equation} \label{x1zVbC}
    \tau_0 
    = 
    \sqrt{2 \gamma_0 \frac{2G}{h}}  ,
\end{equation} 
which gives the strength of the joint in the absence of contamination. In addition, from (\ref{YQ1c5Y}),
\begin{equation} 
    \Delta_\infty
    = 
    \sqrt
    {
        \Big(
            \gamma_0
            -
            \frac{1}{2}
            c_{\rm eq} \Delta g_0 \ell
        \Big)
        \frac{h}{G} 
    }
\end{equation}
follows as the critical displacement for failure when the joint is in equilibrium with an aggressive environment, i.~e., when the joint is loaded after long exposure to the environment so that $c(x,t) \sim c_{\rm eq}$ everywhere in the joint overlap. Using again the relation $\tau \sim 2G \Delta/h$, we find
\begin{equation} \label{rkEwde}
    \tau_\infty
    = 
    \sqrt
    {
        \Big(
            2\gamma_0
            -
            c_{\rm eq} \Delta g_0 \ell
        \Big)
        \frac{2G}{h}
    } ,
\end{equation}
which quantifies the equilibrium strength of the joint. 

We note that, if $\Delta g_0 \geq 0$, i.~e., if the contaminant binds to the adhesive and causes a decrease in its specific fracture energy, then $\tau_\infty < \tau_0$ for $c_{\rm eq} > 0$, which is indicative of a deleterious effect of the environment on the strength of the joint. This type of environmental degradation is clearly evident in test where the joints are exposed for a long time to aggressive conditioning environments (e.~g., \cite{mubashar2009a}). 

%, e.~g., in tensile tests of single lap joints manufactured from aluminium alloy 2024T3 and O and FM73 adhesive carried out after the joints were exposed to different conditioning environments reported by Mubashar {\sl et al.} \cite{mubashar2009a}. The experimental results revealed that the failure strength of the single lap joints with 2024T3 adherends progressively degraded with time when conditioned at 50${}^\circ$C immersed in water. 

When a joint is conditioned in an aggressive environment over a time of exposure shorter than required to attain of equilibrium over the entire joint overlap, the strength of the joint following a subsequent step application of the load is 
\begin{equation} \label{roOGBv}
    \tau(t)
    = 
    \sqrt
    {
        \Big(
            2\gamma_0
            -
            c(0,t) \Delta g_0 \ell
        \Big)
        \frac{2G}{h}
    } ,
\end{equation}
where $t$ is the time of exposure to the environment and $c(0,t)$ is the impurity concentration at the center of the overlap given by the solution (\ref{0vg5Kc}). Using (\ref{PHI89N}) and undoing the normalization (\ref{o8lX7F}), eq.~(\ref{roOGBv}) becomes, explicitly,
\begin{equation} 
    \tau(t)
    = 
    \sqrt
    {
        \Big(
            2\gamma_0
            -
            {\rm erfc}
            \Big(
                \frac{a}{2\sqrt{D t}}
            \Big) 
            c_{\rm eq} \Delta g_0 \ell
        \Big)
        \frac{2G}{h}
    } ,
\end{equation}
Evidently, for $t=0$, $c(0,0) = 0$ and  (\ref{roOGBv}) reduces to $\tau_0$, eq.~(\ref{x1zVbC}). Conversely, for $t\to +\infty$, $c(0,t) \to c_{\rm eq}$ and  (\ref{roOGBv}) reduces to $\tau_\infty$, eq.~(\ref{rkEwde}). 

We thus conclude that $\tau(t)$, the strength of a lap joint exposed to the environment for a time $t$, interpolates smoothly between $\tau_0$ and $\tau_\infty$ on the diffusive time scale. This type of gradual degradation and time scaling is clearly evident in test data \cite{kinloch1979a, gledhill1974a, mubashar2009a, bordes2009a}. 

\section{Summary and outlook} \label{Ow8Ezz}

We have formulated a multiscale model of adhesive layers undergoing impurity-dependent cohesive fracture. The model contemplates three scales: i) at the atomic scale, fracture is controlled by interatomic separation and the thermodynamics of separation depends on temperature and impurity concentration; ii) the mesoscale is characterized by the collective response of a large number of interatomic planes across the adhesive layer, resulting in a thickness-dependence strength; in addition, impurities are ingressed from the environment and diffuse through the adhesive layer; and iii) at the macroscale, we focus on lap joints under the action of static loads and aggressive environments. Within this scenario, we have formulated an analytically tractable one-dimensional model and obtained closed form analytical solutions for: the dependence of the adhesive layer strength on thickness; the length of the edge cracks, if any, as a function of time; the lifetime of the joint; and the dependence of the strength of the joint on time of preexposure to the environment. Overall, the theory is found to capture well the experimentally observed trends where available. The closed form solutions supply ready-to-use design formulas that elucidate explicitly the dependence of design metrics of interest, such as the lifetime of the joint, on material and structural parameters. 

%\begin{figure}[ht!]
%\begin{center}
%    \includegraphics[width=0.99\textwidth]{./Figures/BornHaber}
%    \caption{Left: Model used to calculate interfacial fracture energy as a function of impurity coverage \cite{ven2003a}. Right: The Born-Haber cycle used to calculate the interfacial fracture energy of the interface as a function of impurity concentration.} \label{1IT9Xt}
%\end{center}
%\end{figure}

Perhaps most importantly, multiscale analysis has the virtue of identifying the fundamental material parameters that control fracture-dominated failure of adhesive joints as the specific fracture energy $2 \gamma_0$ of the uncontaminated adhesive, its shear stiffness $G$ and the binding energy $\Delta g_0$, representing the affinity of the adhesive for the impurities. These parameters are known for a limited number of combinations of adhesives and contaminants. A problem arises when new adhesive materials are formulated and there is a need to bring them online. Conveniently, the relevant material parameters identified by the multiscale analysis can be calculated, and new material specifications characterized, using atomistic models. Thus, the effect of impurities on the specific fracture energy can be computed, e.~g., by recourse to the Born-Haber cycle \cite{ven2003a, jiang2004a}. 

The strength of a wide variety of metal/adhesive interfaces has been investigated using Density Functional Theory (DFT) and molecular dynamics (MD) models \cite{semoto2011a, nakamura2022a, sumiya2022a, sumiya2022b}. For instance, Zhang \cite{zhang-a}  has proposed a molecular model for Methacrylate-based adhesives (MMA), including two-component methyl methacrylate resin (A/B) and benzoyl peroxide (BPO), with results of actual pull-off tests from steel substrates also reported in the same paper. The polymethyl methacrylate (pMMA) polymer structures are available in the repeat unit list of system libraries, while the polybutyl acrylate (pnBA) polymer structure can be obtained from \cite{urban2018a}. Computational cells can be built at atmospheric pressure and subsequently annealed from 600K to room temperature followed by an NPT-ensemble relaxation to attain the equilibrium density. Bonded and non-bonded interactions between MMA atoms ca be modeled using the OPLSAA force field \cite{jorgensen1996a, jorgensen1988a}. Non-bonded interactions, including Pauli repulsion, van-der-Waals forces and coulombic interactions can be taken into account via the Lennard-Jones potential, while long-range electrostatic interactions can be evaluated using Ewald summation. 

This wealth of atomistic models can be enlisted to generate tables of fundamental material properties such as fracture energy as a function of impurity coverage, to aid in the characterization and evaluation of long term structural performance of adhesive joints under representative operational and environmental conditions. 

\section*{Acknowledgements}

We gratefully acknowledge the support of the US Office of Naval Research through award N62909-24-1-2079.

\begin{appendices}
\section{The effective cohesive behavior of an adhesive layer}
\label{kRMhEb}

\begin{figure}
\begin{center}
    \includegraphics[width=0.69\textwidth]{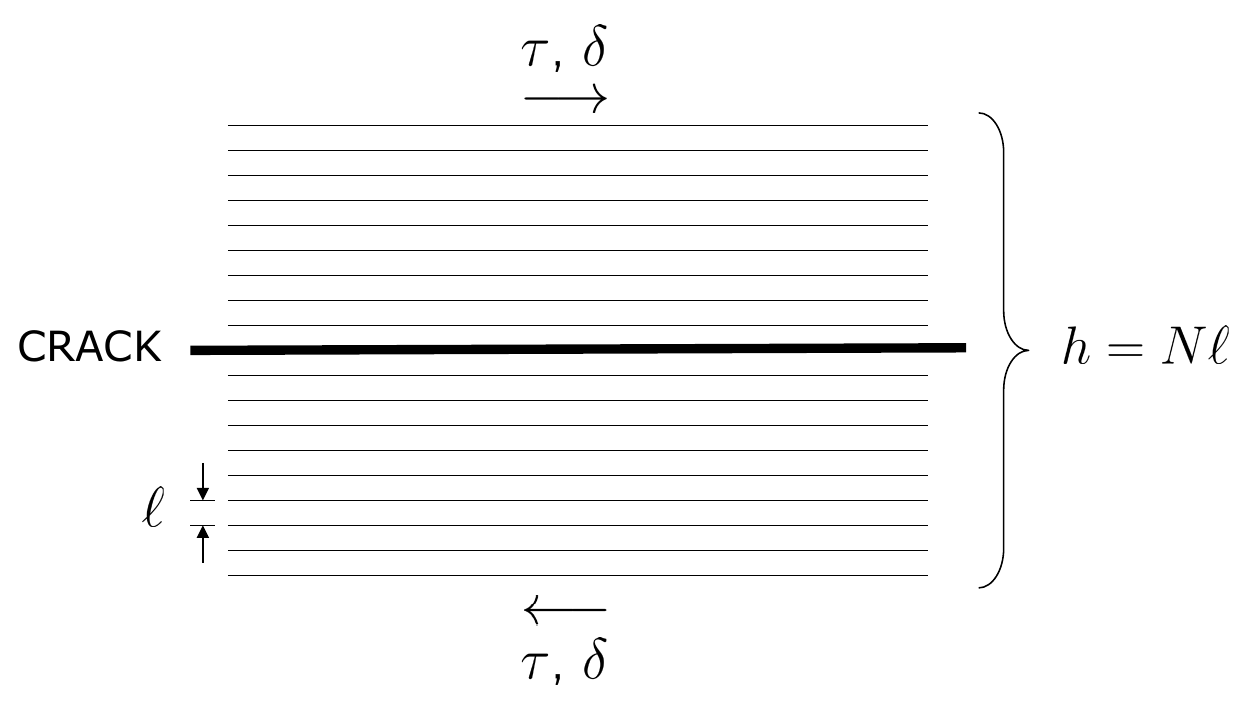}
    \caption{Model of an adhesive layer as a stack of $N$ cohesive planes.} \label{oZNoqI}
\end{center}
\end{figure}

A simple model of cleavage in an adhesive layer consists of a stack of $N$ atomic planes sliding in accordance with an interplanar potential under the action of a sliding displacement $\delta$ prescribed on the boundary of the layer, Fig.~\ref{oZNoqI}. Nguyen and Ortiz \cite{NguyenOrtiz2002} investigated this problem for the special case of nearest-neighbor interactions using formal asymptotics and renormalization group techniques. Braides {\sl et al.} \cite{Braides:2006} extended that work to account for interactions beyond nearest-neighbor planes. 

We summarize the renormalization argument of Nguyen and Ortiz \cite{NguyenOrtiz2002} for completeness. Let $\delta_i \geq 0$, $i = 1,\dots, N$ be the sliding displacements of the interatomic planes in the adhesive layer. Then, the total energy of the adhesive layer is
\begin{equation}\label{Etot}
    E^{\rm tot} = \sum_{i=1}^N  f(\delta_i) ,
\end{equation}
where $f(\delta_i)$ is the interatomic potential. Let now $\delta$ be the macroscopic opening displacement. Then, the effective or macroscopic energy of the cohesive layer follows from the constrained minimization problem
\begin{equation} \label{Energy}
    f_0(\delta) 
    = 
    \inf_{\{\delta_1, \dots, \delta_N\}}
    \sum_{i=1}^N  f(\delta_i),
    \quad
    \delta = \sum_{i=1}^N \delta_i .
\end{equation}
A careful analysis reveals that, asymptotically as $N \to \infty$, multiple decohered planes are not energetically possible favorable, and only the cases of one decohered plane or none need be considered. This alternative give
\begin{equation}
    f_0({\delta}) = \min \{\frac{{C_N}}{2}
    {\delta}^2, 2\gamma \} = \left\{
    \begin{array}{cc}
    ({C_N}/2) {\delta}^2, &
    \text{if } {\delta} < {\delta}_c \\
    2 \gamma, & \text{otherwise}
    \end{array}
    \right.
\end{equation}
where
\begin{equation}
    C = \frac{1}{\ell} (\lambda + 2\mu) ,
    \quad
    {C_N} = \frac{C}{N} ,
    \quad
    {\delta}_c = 2 \sqrt{\frac{\gamma}{{C_N}}} = 2
    \sqrt{\frac{\gamma N}{C}}
\end{equation}
are an interplanar stiffness, an effective cohesive-layer stiffness and a macroscopic critical opening displacement for the nucleation of a single decohered plane, respectively. The corresponding macroscopic cohesive law is
\begin{equation}
    {\tau}({\delta}) = \left\{
    \begin{array}{cc}
    {C_N}{\delta}, &
    \text{if } {\delta} < {\delta}_c \\
    0, & \text{otherwise}
    \end{array}
    \right.
\end{equation}
We note that the effective shear strength of the layer is
\begin{equation}
    {\tau}_c = {C_N} {\delta}_c = 2 \sqrt{{C_N} \gamma}
    = 2 \sqrt{\frac{C \gamma}{N}} .
\end{equation}

Thus, the macroscopic cohesive potential is initially quadratic and subsequently constant following the attainment of the critical macroscopic sliding displacement. Remarkably, the macroscopic critical sliding displacement and shear strength scale as: ${\delta}_c \sim \sqrt{N}$ and ${\tau}_c \sim 1/\sqrt{N}$, respectively. Thus, for large $N$, the adhesive layer sustains much lower shear stresses at much larger sliding displacements than the atomistic binding law. By constrast, the macroscopic fracture energy, or critical energy-release rate, $f_0(\infty)$ remains invariant under the transformation and is equal to the interplanar fracture energy. 

It is also remarkable that the form of the asymptotic form of the interplanar cohesive law is {\sl universal}, i.~e., independent of the atomistic binding law, even when interactions beyond nearest-neighbors are allowed \cite{Braides:2006}. A striking demonstration of universality of the macroscopic cohesive law was supplied by Hayes {\sl et al.} \cite{HayesOrtizCarter2004}, who have computed the macroscopic cohesive law for three different materials and showed that, when the energy and opening displacement are scaled appropriately with respect to $N$, all energy-displacement curves collapse onto a single master curve.

\end{appendices}

\bibliographystyle{sn-mathphys-num}
\bibliography{Biblio,Proposal}

\end{document}